# Triple condensate halo from water droplets impacting on cold surfaces

Yugang Zhao, [1] Fangqi Zhu, [1] Hui Zhang, [1] Chun Yang, [1, *] Tze How New, [1] Liwen Jin [2]

[1] School of Mechanical and Aerospace Engineering, Nanyang Technological University,
50 Nanyang Avenue, Singapore 639798, Singapore

[2] XXX

**Abstract:**

Understanding the dynamics in the deposition of water droplets onto solid surfaces is of importance from both fundamental and practical viewpoints. While the deposition of a water droplet onto a heated surface is extensively studied, the characteristics of depositing a droplet onto a cold surface and the phenomena leading to such behavior remain elusive. Here we report the formation of a triple condensate halo observed during the deposition of a water droplet onto a cold surface, due to the interplay between droplet impact dynamics and vapor diffusion. Two subsequent condensation stages occur during the droplet spreading and cooling processes, engendering this unique condensate halo with three distinctive bands. We further proposed a scaling model to interpret the size of each band, and the model is validated by the experiments of droplets with different impact velocity and varying substrate temperature. Our experimental and theoretical investigation of the droplet impact dynamics and the associated condensation unravels the mass and heat transfer among droplet, vapor and substrate, offer a new sight for designing of heat exchange devices.

## Introduction

Liquid droplets impacting on solid surfaces is ubiquitous in nature and everyday societal and industrial applications. Numerous experimental and theoretical works have been carried out, aiming at describing and predicting the kinetical behavior of impacting droplets[1-9]. Unraveling the physical mechanism pertaining to the impinging, spreading, recoiling/rebounding/splashing processes under different conditions is of significant importance, to promote the performance and safety in a variety of applications from ink-jet printing[10-12], spray cooling[13, 14], self-cleaning surfaces[15-19], to anti-icing in the aviation industry[20-23]. However, factors such as incident velocity, surface topography, thermophysical conditions of the ambient, and liquid properties, make droplets impacting on solid surfaces an intriguing and difficult topic, and remains inadequately understood.

During the impacting of a liquid droplet on a solid surface, mass, momentum and energy are exchanged between droplet and surface. Particularly, such exchange even becomes strong enough to alter the kinetic behavior of droplets, if the surface is superheated ($T_b \gg 100$ ºC) or subcooled ($T_b \ll 0$ ºC). When a droplet impacts on a superheated surface, water evaporates so fast that the droplet floats on its own vapor, referred to as the Leidenfrost effect[24-26]. On the other side, when a droplet impacts on a subcooled surface, the heterogeneous ice nucleation occurs spontaneously at the water/solid interface[27-30]. The contact line is pinned, and thereby the excess kinetic energy is dissipated by constrained trampling instead of spreading[31-34]. Another very common case, where the surface is of a moderate temperature and condensation occurs instead of boiling/icing, has never been studied. Experimental characterization of this case and a through discussion over its mechanism are thus of great fundamental interest.

In this work, we explore in detail the fluid dynamics and thermal dynamics occurring as a warm water droplet impacts on a cold surface ($T_{dp} > T_b > T_{ice}$, with $T_{dp}$ denoting the dew point, $T_{ice}$ the temperature for heterogeneous ice nucleation). We find that a unique triple halo pattern of micrometer-sized condensate droplets is formed during the impact. The formation of such condensate pattern is due to a two-stage condensation, in which the first condensation stage occurs during the droplet spreading process, and the second condensation stage follows until the droplet is cooled down. In addition, we developed a scaling model to elucidate the variation of this halo pattern at different incident velocity and surface temperature, stemming from the interplay between droplet impact dynamics and vapor diffusion.

## Results and discussion

We carry out a direct visualizing experimental study showing what occurs when depositing a water droplet onto a cold surface. Figure 1a shows a simple schematic depiction of the experimental setup. A water droplet (at room temperature of $22 \pm 0.5$ °C) with a fixed volume of $10 \pm 0.2$ μL (typical radius of $r_0 \approx 1.3$ mm) from the quartz capillary is deposited onto the substrate (substrate temperature $T_b = 2.2 \pm 0.1$ °C, $We \approx 6.5$) under gravitational force. The supersaturated vapor from the evaporating droplet condenses and forms isolated micron-sized droplets surrounding the "mother" droplet, referred as a condensate halo. This halo consists three distinctive bands as shown in Figure 1b. Though it is quite rational to anticipate the incidence of condensation when the warm mother droplet approaches the cold surface, a halo with multiple bands is still something intriguing[35].

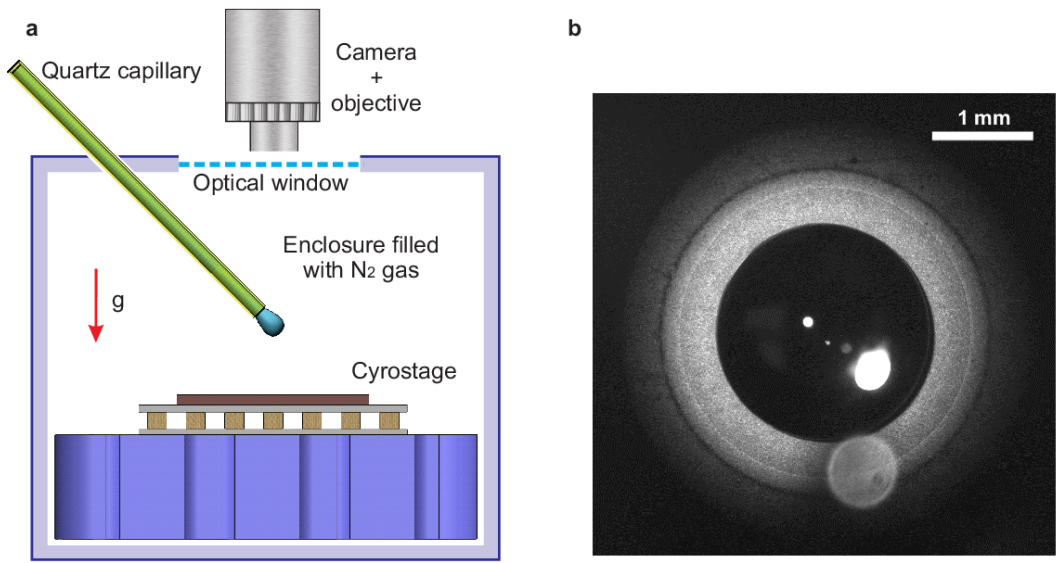

**Figure 1│Triple condensate halo formed when depositing a water droplet onto a cold surface. a**, Schematic depiction of the setup depositing a $10 \pm 0.2$ μL water droplet onto a smooth cold substrate. **b**, A condensate halo with three distinctive bands (at $We \approx 6.5$, and $T_s = 2.2 \pm 0.1$ °C).

Figure 2a shows a segment of the halo, with $R_0$, $R_1$, $R_2$ and $R_3$ being the base radius of the droplet and the locations of three subsequent boundaries along the radical direction. As the third band is relative darker, the inlet shows its location by enhancing the local brightness/contrast. The width of three bands is thus represented by $|R_1 - R_0|$, $|R_2 - R_1|$ and $|R_3 - R_2|$, respectively. The size of each band can be readily detected based the sharp changes of intensity as shown in Supplementary Figure S5. As the halo is composed of micron-sized condensate droplets, each band intrinsically represents the variation of condensed droplet size and number density. Particularly, the boundaries indicate the locations where drastic changes of condensation occur. By calibrating the correlation between the intensity and condensate density (mass per unit area), we can withdraw the information of condensation along radical direction. We used dropwise condensation on the same surface at $RH = 40 \pm 5$ and $T_s = 2.2 \pm 0.1$ °C for the calibration. The condensate density can be computed by using an optical microscope and an image analyzing software to count droplet size and number density. A simple power law fitting formula $y = (x/x_{ref})^\beta$ using moving least square method is shown in Supplementary Figure S5, where $y$ is the normalized intensity, $x$ is local condensate density, $x_{ref}$ is the reference condensate density (~ 305 mg/cm$^2$), and $\beta$ (~ 0.47) is the power constant. Based on this calibration result, the variation of condensate density along the radical direction is shown Figure 2b. The location of $R_1$, $R_2$ and $R_3$ can thereby be found at the places where the condensate density drastically increases/decreases, depending on the nature of each band.

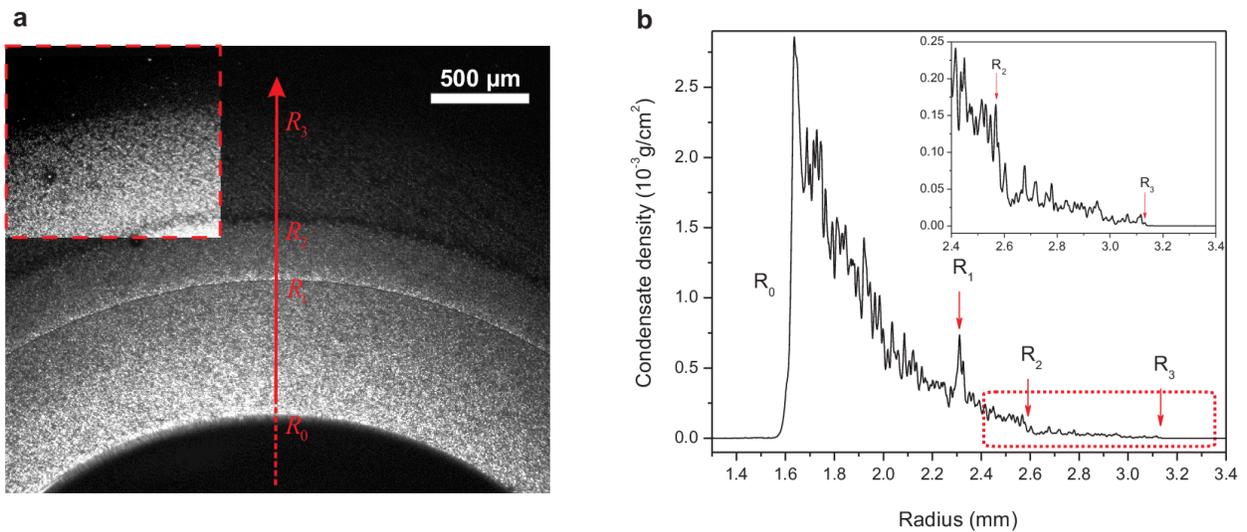

**Figure 2 | Segment of the condensate halo showing the width of each band. a**, an optical microscope image showing the variation of intensity along the radical direction (at $We \approx 6.5$, and $T_b = 2.2 \pm 0.1$ °C). Inlet shows the boundary of band $|R_3 - R_2|$ by enhancing the local

brightness/contrast. **b**, Variation of computed condensate density along the radical direction based on the optical image of **a**, with $R_0$, $R_1$, $R_2$ and $R_3$ being the locations where drastic changes occur. Inlet shows a magnified plot in the selected rectangle.

To achieve further insight of the mechanism how the condensate halo is formed, we employed two synchronized cameras to observe the droplet impact dynamics and the growth of condensate halo from both side view and top view. Figure 3a shows the impact dynamics of a water droplet onto a cold surface from side view, with $t = 0$ ms being the moment the droplet contacts the substrate. At $t \approx 10$ ms, the dynamic radius $R$ approaches the maximal spreading[36, 37] of $R_{max} \sim r_0 We^{1/4}$, and the contact line starts to retract. A stable droplet showing the equilibrium contact angle $\theta$ and base radius $R_0$ is found at $t \approx 1$ s. Figure 3b shows schematically how the condensate halo grows as the droplet impacts onto the cold surface. The pendent droplet is surrounded by its own vapor before its falling upon gravitational force, due to the diffusion controlled evaporation. The vapor pressure is undersaturated at in-chamber temperature $T_\infty$, but supersaturated at substrate temperature $T_b$. Thereby, the excess vapor is condensed onto the cold surface and forms the condensate halo. Nevertheless, instead of one continuous halo, a condensate halo composed of three distinctive bands is formed due to the interplay between droplet spreading dynamics and vapor diffusion. Condensation occurs in two subsequent stages. The first stage condensation occurs during the spreading process, and condensate droplets only distribute from $R_{max}$ to certain distance outwards along the radical direction (otherwise, being removed by the advancing contact line). The width of this band $|R_2 - R_1|$ is thus evaluated by the characteristic length of vapor diffusion within the first stage $l_s \sim \sqrt{Dt_s}$, with $D$ being the diffusivity of water vapor, and $t_s$ the spreading time. $t_s$ is independent of the impact velocity [38] and scales as $\mu r/\sigma$, where $\mu$ and $\sigma$ are dynamic viscosity of the fluid and surface tension of the free interface, respectively. The second stage condensation occurs afterwards till the droplet is cooled and evaporation is suppressed, thus feeding vapor flux from the evaporation of mother droplet can't support further condensation at $R > R_3$. The width of $|R_3 - R_0|$ is evaluated by the characteristic length of vapor diffusion within the second stage $l_c \sim \sqrt{Dt_c}$, with $t_c$ being the characteristic time of droplet cooling. A direct energy balance analysis shows $t_c$ scales as $R_0^2 f_1(\theta) \frac{1}{\alpha}$, where $f_1(\theta) = \left(\dfrac{\sin\theta}{1+\cos\theta}\right)^3 \left(1 - \dfrac{\sin\theta}{3+3\cos\theta}\right)$ is a geometrical factor, and $\alpha$ is thermal diffusivity of the fluid. The derivation detail of $t_c$ can be found in supplementary information. Figure 3c shows the growth dynamics of condensate halo from top view. The results show clearly that $R_1$ coincides with $R_{max}$, and the band $|R_2 - R_1|$ emerges right after the spreading process.

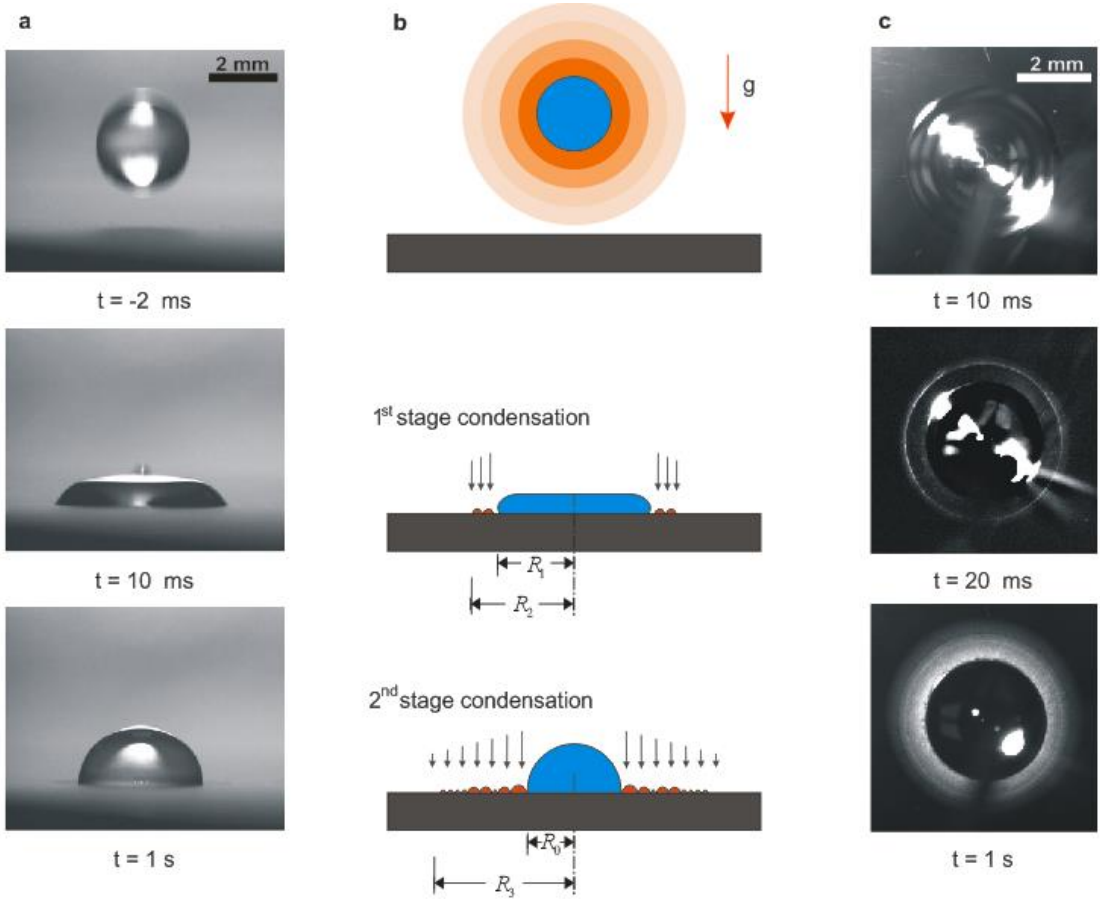

**Figure 3 | Droplet impact dynamics and growth of condensate halo. a**, Optical images (side view) showing the impact dynamics of a droplet (at $We \approx 6.5$, and $T_b = 2.2 \pm 0.1$ ºC). **b**, Schematics showing the growth of condensate halo due to the dual-stage condensation. **c**, Corresponding optical images (top view) from **a** showing the growth dynamics of condensate halo.

Some key features of the condensate halo can be anticipated based on our understanding of its physics. For example, we can evaluate the spreading time $t_s \sim 10$ ms experimentally, as showing in supporting movie 1, 2 and Figure 3. Also, the characteristic time of droplet cooling $t_c \sim 1$ s can be evaluated from the thermal measurement as showing in supporting Figure S3. Thereby, it gives $|R_2 - R_1|/|R_3 - R_0| \sim l_s / l_c \sim 10$, which matches the experimental results. Figure 4a shows the variation of each band width as a function of the impact velocity (*We*). For a given droplet volume and the surface chemistry (*r*, $R_0$, *θ*), the dynamic local vapor concentration is solely determined by the material properties, and independent of the impact velocity. Both of two characteristic lengths, $|R_2 - R_1|$ and $|R_3 - R_0|$ are constant. However, as $R_1 = R_{max}$ scales as $rWe^{1/4}$, band_1 of $|R_1 - R_0|$ increases and band_3 of $|R_3 - R_2|$ decreases with increasing impact velocity. A special case occurs at *We* < 1, i.e. the droplet is gently deposited on to the substrate, and thus it gives $R_2 \approx R_1 \approx R_0$. Figure 4b shows the variation of each band width as a function of substrate temperature. As the temperature dependence of material properties (*D*, *μ*, *σ* and *α*) in this work is negligible, the dynamic local vapor concentration is independent of the substrate temperature. However, the degree of supersaturation $s \equiv RH(T_\infty)/RH(T_b)$ increases drastically with decreasing substrate temperature, as shown in supplementary information. Accordingly, the energy barrier of heterogeneous condensation nucleation $\Delta G = \dfrac{16\pi M^2 \sigma^3}{3(RT\rho \ln S)^2} f_2(\theta)$ decreases with decreasing substrate temperature, where $f_2(\theta) = \dfrac{1}{2} - \dfrac{3}{4}\cos\theta + \dfrac{1}{4}\cos^3\theta$ is also a surface wetting factor, *M* is the molecular weight of liquid, *R* is idea gas constant, and *ρ* is the density of liquid. Both of two characteristic lengths, $|R_2 - R_1|$ and $|R_3 - R_0|$ increase with decreasing substrate temperature, while the change of band_1 $|R_1 - R_0|$ is negligible.

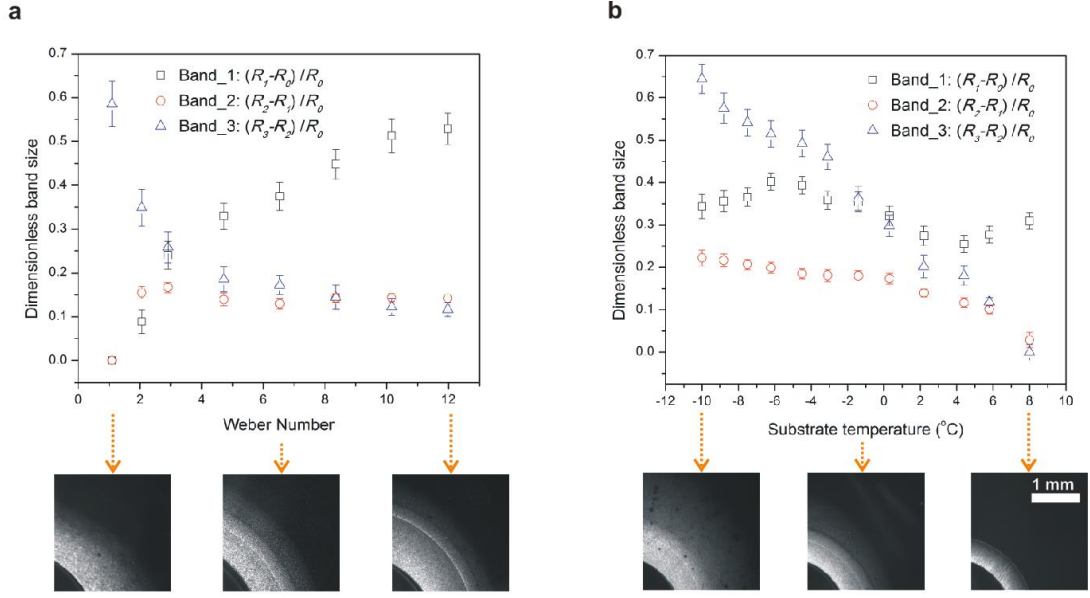

**Figure 4 │ a**, Variation of the band width as a function of substrate temperature (at $T_s = 2.2 \pm 0.1$ °C). **b**, Variation of the band width as a function of substrate temperature (at $We \approx 6.5$).

## Conclusion

In summary, we reported the formation of a triple condensate halo when depositing a water droplet onto a cold surface. We have shown both experimentally and theoretically that this condensate halo is formed due to the interplay between droplet impact dynamics and vapor diffusion. Two subsequent condensation stages give the condensate halo a unique feature of three bands with distinctive boundaries, allowing to evaluate the size of each band basing our understanding of the mechanism. From a boarder perspective, this work represents an important advance in our understanding of the droplet impacting with phase changes. We envisage that such an intriguing condensation behavior will find a range of applications in highly demanding heat transfer devices.

## Methods

**Sample surfaces.**

Single-side polished single crystal silicon wafers were used as substrates in this study due to their atomic smoothness and dark background. Each wafer was cut into square pieces of 2 cm × 2 cm in dimension with a diamond cutter. Substrates were then cleaned using a standard protocol, which involves rinsing with acetone, isopropyl alcohol, ethanol, and deionized (DI) water subsequently and drying in nitrogen gas flow. The surface energy of substrates was modified through chemical vapor deposition of

trichloro(1H,1H,2H,2H-perfluorooctyl)-silane monolayer (Sigma Aldrich) right after an oxygen plasma treatment of 30 mins. The wettability of substrates was characterized by measuring their static contact angles (Attension Theta), and the results are $\theta = 91.5 \pm 2°$. Right before the test, substrates were rinsed again with DI water for 1 mins to remove the deposited dust from ambient air. Afterwards, substrates were bonded onto the cryostage with a thin layer (approx. 100 μm) of silicone grease (Omega Engineering) to minimize thermal contact resistance.

**Experimental apparatus.**

Our experimental apparatus is an enclosure chamber with a cryostage inside as shown in Supplementary Figure S1, allowing for directly observing the droplet impact dynamics and the formation of condensate halo. The chamber was made of Teflon frame with acrylic windows and filled with nitrogen gas with a dry-bulb temperature as $22 \pm 0.5$ °C. In-chamber pressure was well balanced with the ambient pressure via a one-way check valve. The cryostage consisted of a heat sink and a Peltier element (Ferrotec, 9500/127/060B) to attain dual-stage control of substrate temperature. The substrate temperature was maintained from -10 °C to 8.0 °C using a proportional-integral-derivative controller via a LabView data acquisition input & output module (National Instrument, NI 9219 & NI 9264). Silicone grease was applied between the heat sink and the Peltier element to minimize thermal contact resistance. During experiments, single liquid droplet was generated by a quartz capillary (400μm ID and 500μm OD) that was connected to a syringe pump operating at a constant flowrate of 100 μL/min. The droplet had a volume of $10 \pm 0.2$ μL (with a typical radius of $r_0 \approx 1.3$ mm), and it was released on the center area of the subcooled substrate at a certain height with Weber number varying from 1.1 to 12.0. Here $We = 2\rho g h r_0 / \mu$, where $\rho$ is the density of fluid, $g$ is the gravitational acceleration, $h$ is the deposition height and $\mu$ is the fluid dynamic viscosity. Three T-type thermocouples were applied to monitor the in-chamber, droplet, and substrate temperatures, respectively. The alignment of thermocouples is shown in Supplementary Figure S2. Two synchronized cameras were employed, with one high speed (Phantom. M310, 1000 FPS) of a side view and one high quantum efficiency (PCO. edge 4.2, 100 FPS) of a top view, to capture the droplet impact dynamics and growth dynamics of the condensate halo, respectively.

**Image processing and data acquisition.**

To gain the detailed information of condensate halo, the captured images from Camera 1 as shown in Supplementary Figure S1 was analyzed using a Matlab-based image processing approach. An example is provided in supplementary Figure S4. The original image was first subtracted of a background (the image before droplet emerges). Noise suppression was applied using a Savitzky-Golay method. The grey value of the image was then normalized and plotted along the radical direction of the droplet.